\documentclass[fleqn,usenatbib]{mnras}

\usepackage{newtxtext,newtxmath}

\usepackage[T1]{fontenc}

\DeclareRobustCommand{\VAN}[3]{#2}
\let\VANthebibliography\thebibliography
\def\thebibliography{\DeclareRobustCommand{\VAN}[3]{##3}\VANthebibliography}

\usepackage{graphicx}	
\usepackage{amsmath}	
\usepackage{mathtools}

\title[Collective tidal mode damping]{
On the damping of tidally driven oscillations
}

\author[J. W. Dewberry \& S. C. Wu]{
Janosz W. Dewberry,$^{1}$\thanks{E-mail: jdewberry@cita.utoronto.ca}
Samantha C. Wu,$^{2}$
\\
$^{1}$CITA, 60 St. George Street, Toronto, ON M5S 3H8, Canada\\
$^{2}$TAPIR, Walter Burke Institute for Theoretical Physics, Mailcode 350-17, California Institute of Technology, Pasadena, CA 91125, USA
}

\date{Accepted XXX. Received YYY; in original form ZZZ}

\pubyear{2023}

\begin{document}
\label{firstpage}
\pagerange{\pageref{firstpage}--\pageref{lastpage}}
\maketitle

\begin{abstract}
Expansions in the oscillation modes of tidally perturbed bodies provide a useful framework for representing tidally induced flows. However, recent work has demonstrated that such expansions produce inaccurate predictions for secular orbital evolution when mode damping rates are computed independently. We explore the coupling of collectively driven modes by frictional and viscous dissipation, in tidally perturbed bodies that are both non-rotating and rigidly rotating. This exploration leads us to propose an alternative approach to treating the damping of tidally driven oscillations that accounts for dissipative mode coupling, but which does not require any information beyond the eigenfunctions and eigenfrequencies of adiabatic modes.
\end{abstract}



\section{Introduction}
Tides alter the orbital structures of a wide range of astrophysical systems \citep{Ogilvie2014}. The difficulty of placing quantitative constraints on this influence, in particular the transfer of energy and angular momentum between tidally interacting bodies and their orbits, motivates exploration of any viable methods for computing tidal dissipation. One approach involves describing the tidal flow raised in a tidally perturbed planet, star or compact object in terms of that body's oscillation modes, and has remained popular for several decades \citep[e.g.,][]{Press1977,Kumar1995,Schenk2001,Wu2005a,Wu2005b,Burkart2012,Braviner2015,Fuller2017,Xu2017,Yu2021,Dewberry2022a}. Mode expansions are advantageous because they often provide a sparse representation of the tidal flow, and because the characteristics of the most strongly driven modes in a given tidal interaction can elucidate the physics at play. 

Recent work \citep{Sun2023,Townsend2023} has undermined these advantages by showing that the tidal torques estimated from the simplest approach to mode expansions can deviate significantly from the results of direct, mode-independent solution of the (Fourier-transformed) governing equations \citep[e.g.][]{Ogilvie2009,Ogilvie2013}. \citet{Townsend2023} argue that this discrepancy, which can be significant enough to produce diverging predictions for secular orbital evolution, originates in the assumption that mode damping can be treated individually for each mode \citep[the fallacy of this assumption is also evident in the analysis of ][]{Braviner2015}. \citet{Townsend2023} further suggest that the disagreement can be reconciled with the use of a universal, tidal frequency-dependent damping rate that agrees with individual mode damping rates only at exact resonance. However, the authors' construction of such a universal damping rate relies on information from direct tidal calculations that mode expansions are intended to replace, limiting its utility.

In this paper, we expand upon the analyses of \citet{Braviner2015} and \citet{Townsend2023}, and suggest an alternative approach to treating mode damping that relies only on information contained in the eigenfunctions of adiabatic oscillation modes. Working directly from the equation of motion, we explore some of the ways in which coupling between mode amplitudes can depend on the form of damping included. We then show that this coupling can be bypassed by making use of a fundamental relationship \citep{Ogilvie2013} between the total dissipation rate and the imaginary parts of tidal Love numbers. We focus on non-rotating bodies in \autoref{sec:nonrot}, provide a generalization for rigidly rotating bodies in \autoref{sec:rot}, and conclude in \autoref{sec:conc}. 

\section{Non-rotating bodies}\label{sec:nonrot}
The equation of motion for the Lagrangian displacement $\boldsymbol{\xi}$ induced in a non-rotating fluid body by a tidal interaction can be written as
\begin{equation}\label{eq:tEoM}
    \frac{\partial^2\boldsymbol{\xi}}{\partial t^2}
    +{\bf C}[\boldsymbol{\xi}]
    +{\bf D}[\boldsymbol{\xi}]
    ={\bf f},
\end{equation}
where ${\bf C}$ is a self-adjoint linear operator, ${\bf D}$ is an operator that describes dissipation (which we assume to be linear), 
and ${\bf f}$ is the tidal force per unit mass. In this paper we assume the form ${\bf f}=-\nabla U$ for some potential $U\propto\exp[-\text{i}\omega_tt]$ with a harmonic dependence on a real-valued frequency $\omega_t$ (throughout, the physical displacement and tidal force should be taken as the real parts of $\boldsymbol{\xi}$ and ${\bf f}$). In the absence of rotation in  the tidally perturbed body, it is useful to introduce an expansion  $\boldsymbol{\xi}({\bf r},t)=\sum_\beta a_\beta(t)\boldsymbol{\xi}_\beta({\bf r})$ in the distinct eigenfunctions of oscillation modes with displacements $\hat{\boldsymbol{\xi}}_\beta({\bf r},t)
=\boldsymbol{\xi}_\beta({\bf r})\exp[-\text{i}\omega_\beta t]$ normalized to satisfy
\begin{align}
    {\bf C}[\boldsymbol{\xi}_\beta]
    &=\omega_\beta^2\boldsymbol{\xi}_\beta,
\\
    \langle 
        \boldsymbol{\xi}_\alpha,\boldsymbol{\xi}_\beta
    \rangle
    &=\int_V\rho_0
    \boldsymbol{\xi}_\alpha^*\cdot\boldsymbol{\xi}_\beta\text{d}V
    =MR^2\delta_{\alpha\beta},
\end{align}
where $\rho_0(r)$ is the equilibrium density of the tidally perturbed body, $M$ is its total mass, and $R$ is its radius. Inserting this expansion into the equation of motion and taking the inner product with the eigenfunction $\boldsymbol{\xi}_\alpha$ of a given mode $\alpha$ then leads directly to the amplitude equation
\begin{equation}\label{eq:amp1}
    \ddot{a}_\alpha
    +\omega_\alpha^2a_\alpha
    +\sum_\beta\langle\boldsymbol{\xi}_\alpha,{\bf D}[a_\beta\boldsymbol{\xi}_\beta]\rangle
    =-\langle\boldsymbol{\xi}_\alpha,\nabla U\rangle.
\end{equation}
Decoupling of this equation between different driven oscillation mode amplitudes then depends on the orthogonality of a given $\boldsymbol{\xi}_\alpha$ with ${\bf D}[a_\beta\boldsymbol{\xi}_\beta]$ for $\beta\not=\alpha.$ 

\subsection{Frictional damping}
The simplest case to consider is that of a frictional (Stokes) damping with ${\bf D}[\boldsymbol{\xi}]=2\gamma {\bf v},$
where $\gamma$ is a constant and
\begin{equation}
    {\bf v}=\partial_t\boldsymbol{\xi}
    =\sum_\beta \dot{a}_\beta \boldsymbol{\xi}_\beta
\end{equation}
is the velocity field associated with the tide. For such dissipation, the amplitude equations separate into a set of decoupled driven, damped harmonic oscillator equations for each mode amplitude
\begin{equation}
    \ddot{a}_\alpha
    +2\gamma\dot{a}_\alpha
    +\omega_\alpha^2a_\alpha
    =-\langle\boldsymbol{\xi}_\alpha,\nabla U\rangle.
\end{equation}
This equation has the steady-state ($\dot{a}_\alpha=-\text{i}\omega_ta_\alpha$) solution
\begin{equation}\label{eq:aafric}
    a_\alpha 
    =\frac{-\langle\boldsymbol{\xi}_\alpha,\nabla U\rangle}
    {(\omega_\alpha^2 - \omega_t^2) -2\text{i}\gamma\omega_t}.
\end{equation}
The amplitude equations separate in this special case because the eigenfunctions of the inviscid oscillation equations (i) happen to be eigenfunctions of the dissipative operator (since it simply involves multiplication by the constant $-2\text{i}\omega_t\gamma$), and are (ii) orthogonal under the inner product $\langle\ ,\rangle$. Whenever either of these conditions are not satisfied, the amplitudes of the collectively driven modes do not individually satisfy \emph{decoupled} harmonic oscillator equations.

\subsection{Viscous damping}
As an alternative example, consider viscous dissipation:
\begin{equation}
    {\bf D}[\boldsymbol{\xi}]
    =-\frac{1}{\rho_0}\nabla\cdot(2\mu\delta {\bf S}),
\end{equation}
where $\mu$ is a dynamic viscosity, 
\begin{align}
    \delta {\bf S}
    &=\frac{1}{2}\left[
        \nabla{\bf v}
        +(\nabla{\bf v})^T
        -\frac{2}{3}(\nabla\cdot{\bf v}){\bf I}
    \right]
\\\notag
    &=\omega_t\sum_\beta 
    \frac{a_\beta}{\omega_\beta}
    \underbrace{\frac{1}{2}\left[
        \nabla{\bf v}_\beta
        +(\nabla{\bf v}_\beta)^T
        -\frac{2}{3}(\nabla\cdot{\bf v}_\beta){\bf I}
    \right]
    }_{\textstyle
    \coloneqq\delta{\bf S}_\beta},
\end{align}
and ${\bf v}_\beta=-\text{i}\omega_\beta\boldsymbol{\xi}_\beta$ is the velocity eigenfunction of the mode labelled by $\beta$. Note that with the mode expansion used in this section, the velocity fields of the eigenmodes and the tide are related by 
${\bf v}=\sum_\beta (\omega_t/\omega_\beta)
a_\beta {\bf v}_\beta]
\not=\sum_\beta a_\beta{\bf v}_\beta$. This non-intuitive relationship between tidal and mode velocities results from the choice (in this section) to expand only the spatial part of the tidal displacement in terms of the spatial eigenfunctions of the modes; the modes' frequencies play no role in this expansion, appearing in the amplitude equation only because of the eigenvalue problem satisfied by the oscillations.

\begin{figure*}
    \centering
    \includegraphics[width=\textwidth]{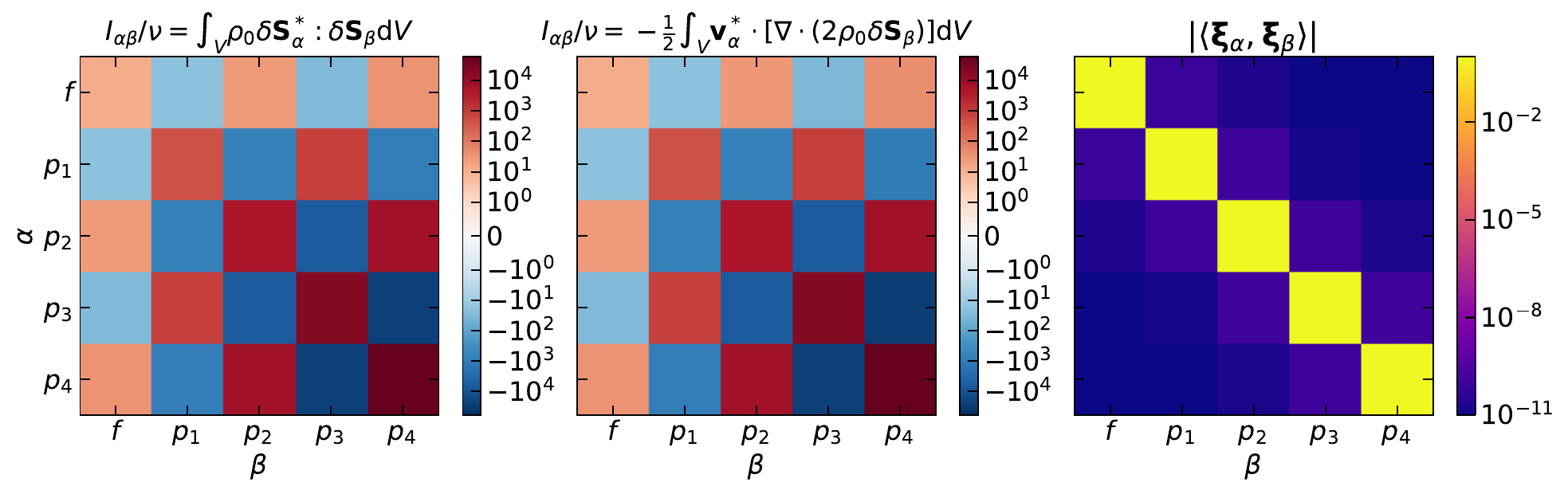}
    \caption{Left, middle: separate calculations of the dissipative cross-coupling integrals given by \autoref{eq:Iab} (normalized by a constant kinematic viscosity $\nu=\mu/\rho_0$), computed for the $\ell=2$ f-mode and several p-modes of an isentropic, non-rotating, $n=1$ polytrope. The colormaps transition from log to linear scale at $I_{\alpha\beta}/\nu=1$. Right: inner products of the same mode eigenfunctions computed with the numerical quadrature used for the left and middle panels. This figure demonstrates that the integrals $I_{\alpha\beta}$ are nonzero for $\alpha\not=\beta$, to a degree more than eleven orders of magnitude larger than the error of the numerical integration.}\label{fig:int}
\end{figure*}

The standard ``quasi-adiabatic'' approach \citep[e.g.,][]{Kumar1995,Burkart2012}
to such dissipation involves replacing the constant $\gamma$ in \autoref{eq:aafric} with individual mode damping rates given by \citep[e.g.,][]{Ipser1991} $\gamma_\alpha=I_{\alpha\alpha}/(\omega_\alpha^2\langle
\boldsymbol{\xi}_\alpha,
\boldsymbol{\xi}_\alpha
\rangle)$, where
\begin{equation}\label{eq:Iab}
    I_{\alpha\beta}
    =-\frac{1}{2}\int_V{\bf v}_\alpha^*\cdot[\nabla\cdot(2\mu\delta {\bf S}_\beta)]\text{d}V
    =\int_V\mu (\delta {\bf S}_\alpha^*:\delta {\bf S}_\beta)\text{d}V.
\end{equation}
The assumption that each mode is damped at all tidal frequencies by its own $\gamma_\alpha$ is only valid if $I_{\alpha\beta}=0$ for $\beta\not=\alpha$, though, and \autoref{fig:int} demonstrates that this is far from guaranteed. For the ($\ell=m=2$) f-mode and lowest radial order p-modes of an $n=1,$ nonrotating and isentropic polytrope, the left and center panels show that the cross-integrals $I_{\alpha\beta}$ instead take values many orders of magnitude larger than the error associated with the numerical (Clenshaw-Curtis) quadrature used (the ``orthonormality'' matrix shown in the right panel demonstrates the bounds of this error).

\begin{figure*}
    \centering
    \includegraphics[width=\textwidth]{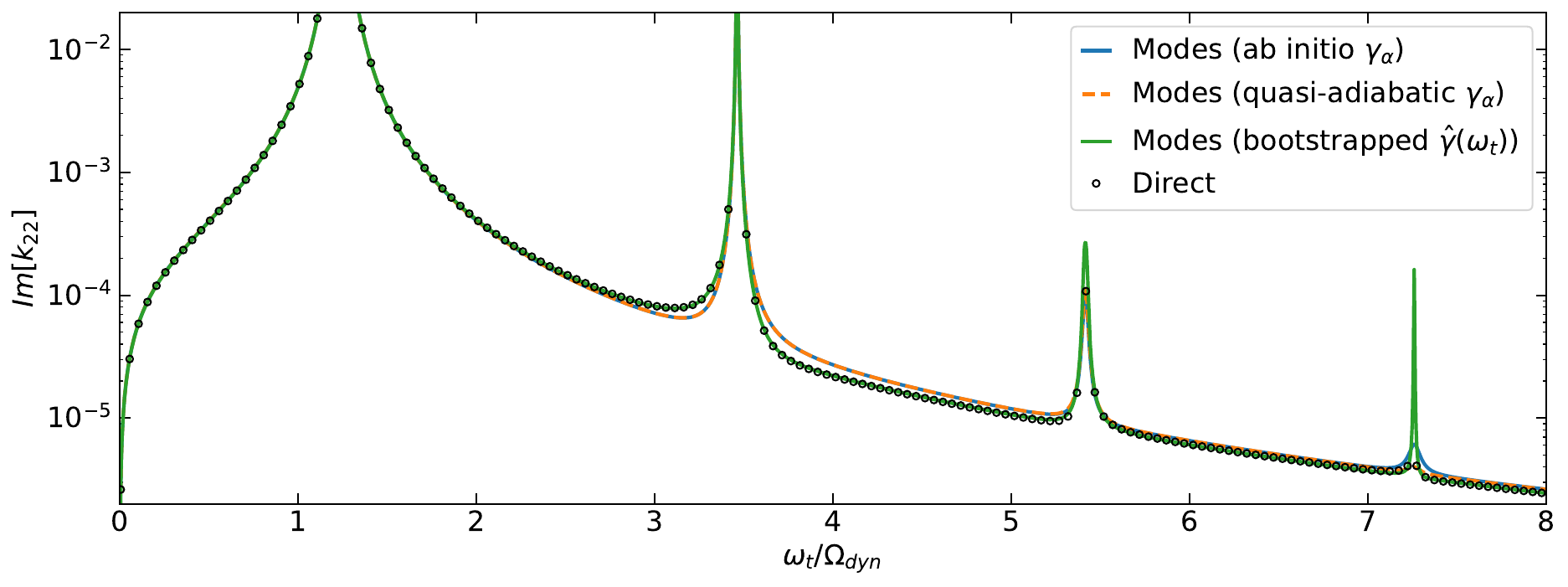}
    \caption{Separate calculations of the imaginary part of the Love number $k_{22}$ for a non-rotating, $n=1$ polytrope with constant kinematic viscosity $\nu=\mu/\rho_0=10^{-4}R^2\Omega_\text{dyn}$. The blue and orange lines plot profiles of $\Im[k_{22}]$ computed using a mode expansion with individual mode damping rates (respectively determined as the imaginary parts of modes computed from the viscous oscillation equations, and through a quasi-adiabatic treatment of inviscid modes). Away from resonances with f-modes and p-modes, these profiles disagree with direct solutions of the governing PDEs at fixed tidal frequency $\omega_t$ (black dots). On the other hand, a mode expansion employing a universal but frequency-dependent damping rate (\autoref{eq:gmc}) agrees well with the direct calculations.}
    \label{fig:visc_nonrot}
\end{figure*}

In the case of viscous dissipation, therefore, direct projection of \autoref{eq:amp1} does not lead to decoupled amplitude equations with individual mode damping rates. \citet{Braviner2015} noted this coupling between modes by viscosity in their analysis of tidal oscillation driving in incompressible Maclaurin spheroids (see their eq. 9). We consequently find, like \citet{Townsend2023}, that mode expansions of the tidal response involving individual mode damping rates inserted into decoupled harmonic oscillator equations disagree with direct numerical calculations of the total tidal response (see \autoref{fig:visc_nonrot}). 

However, we do find that it is possible to construct a universal but \emph{frequency-dependent} damping $\hat{\gamma}=\hat{\gamma}(\omega_t)$ such that $\sum_\beta\langle \boldsymbol{\xi}_\alpha,{\bf D}[a_\beta\boldsymbol{\xi}_\beta]\rangle=2\hat{\gamma}(\omega_t)\dot{a}_\alpha,$ using only the information carried by the adiabatic oscillation modes of the tidally perturbed body. First note that the energy dissipation per unit volume in a viscous fluid is given by $2\mu \Re[\delta {\bf S}]:\Re[\delta {\bf S}]$ \citep[e.g.,][]{Batchelor2000}. The total time-averaged dissipation due to the tidal perturbation of a viscous body is then given by
\begin{align}\label{eq:D}
    \mathcal{D}
    &=\frac{1}{2}\Re\langle {\bf v},{\bf D}[\boldsymbol{\xi}]\rangle
    =\Re\left(
        \omega_t^2\sum_{\alpha,\beta}
        \frac{a_\alpha^*a_\beta}{\omega_\alpha^*\omega_\beta}
        I_{\alpha\beta}
    \right).
\end{align}
If this dissipation is driven by a tidal potential with the form $U=A(r/R)^\ell Y_\ell^m(\theta,\phi)\exp[-\text{i}\omega_t t]$, then it can be related to the imaginary part of a Love number $k_{\ell m}=B/A,$ where $B$ is the harmonic coefficient in the gravitational response 
$\delta\Phi=B(R/r)^{\ell+1}Y_\ell^m(\theta,\phi)\exp[-\text{i}\omega_t t]$ of the same degree and azimuthal order \citep{Ogilvie2014}:
\begin{equation}\label{eq:diss}
    \mathcal{D}=\frac{(2\ell+1)}{8\pi G}R|A|^2\omega_t\Im[k_{\ell m}],
\end{equation}
where $G$ is the universal gravitational constant. Assuming from the outset that $\sum_\beta\langle \boldsymbol{\xi}_\alpha,{\bf D}[a_\beta\boldsymbol{\xi}_\beta]\rangle=2\hat{\gamma}(\omega_t)\dot{a}_\alpha$ for some frequency-dependent $\hat{\gamma}(\omega_t)$, and working in units with $G=M=R=1$, the Love number $k_{\ell m}$ can in turn be found from
\begin{equation}\label{eq:lovenr}
    k_{\ell m}
    =\sum_\beta a_\beta \left(\frac{B_\beta}{A}\right)
    =\frac{4\pi}{(2\ell+1)}\sum_\beta\frac{|Q_{\ell m}^\beta|^2}
    {(\omega_\beta^2 - \omega_t^2) -2\text{i}\hat{\gamma}\omega_t},
\end{equation}
where $B_\beta$ is the harmonic coefficient of the mode $\beta$'s ($\ell,m$) external gravitational perturbation (expressed in solid harmonics associated with spherical coordinates), and 
\begin{equation}
    Q_{\ell m}^\beta=\langle 
        \boldsymbol{\xi}_\beta,\nabla(r^\ell Y_\ell^m)
    \rangle
    =-\frac{(2\ell+1)}{4\pi}B_\beta
\end{equation}
are coefficients describing each mode's overlap with the tidal force. Inserting \autoref{eq:lovenr} into \autoref{eq:diss}, we find
\begin{align}
    \frac{\mathcal{D}}{|A|^2}
    &=
     \sum_\beta\frac{|Q_{\ell m}^\beta|^2\hat{\gamma}\omega_t^2}
    {(\omega_\beta^2 - \omega_t^2)^2}
    \left[
        1
        +\left(
            \frac{2\hat{\gamma}\omega_t}{\omega_\beta^2 - \omega_t^2}
        \right)^2
    \right]^{-1}
.
\end{align}
Then to leading order in the parameter $\hat{\gamma}\omega_t/(\omega_\beta^2 - \omega_t^2)$ (which is small far from resonance with a given mode $\beta$), the frequency dependent, universal damping rate $\hat{\gamma}$ can be estimated from
\begin{align}\label{eq:gmc}
    \hat{\gamma}
    &\simeq 
    \left(    
        \sum_\alpha\frac{|Q_{\ell m}^\alpha|^2\omega_t^2}
        {(\omega_\alpha^2 - \omega_t^2)^2}
    \right)^{-1}\frac{\mathcal{D}}{|A|^2}
\\\notag&
    =
    \left(    
        \sum_\alpha\frac{|Q_{\ell m}^\alpha|^2}
        {(\omega_\alpha^2 - \omega_t^2)^2}
    \right)^{-1}
    \Re\left(\sum_{\alpha,\beta}
    \frac{a_\alpha^*a_\beta I_{\alpha\beta}}{|A|^2\omega_\alpha^*\omega_\beta}
    \right)
    .
\end{align}

\begin{figure*}
    \centering
    \includegraphics[width=\textwidth]{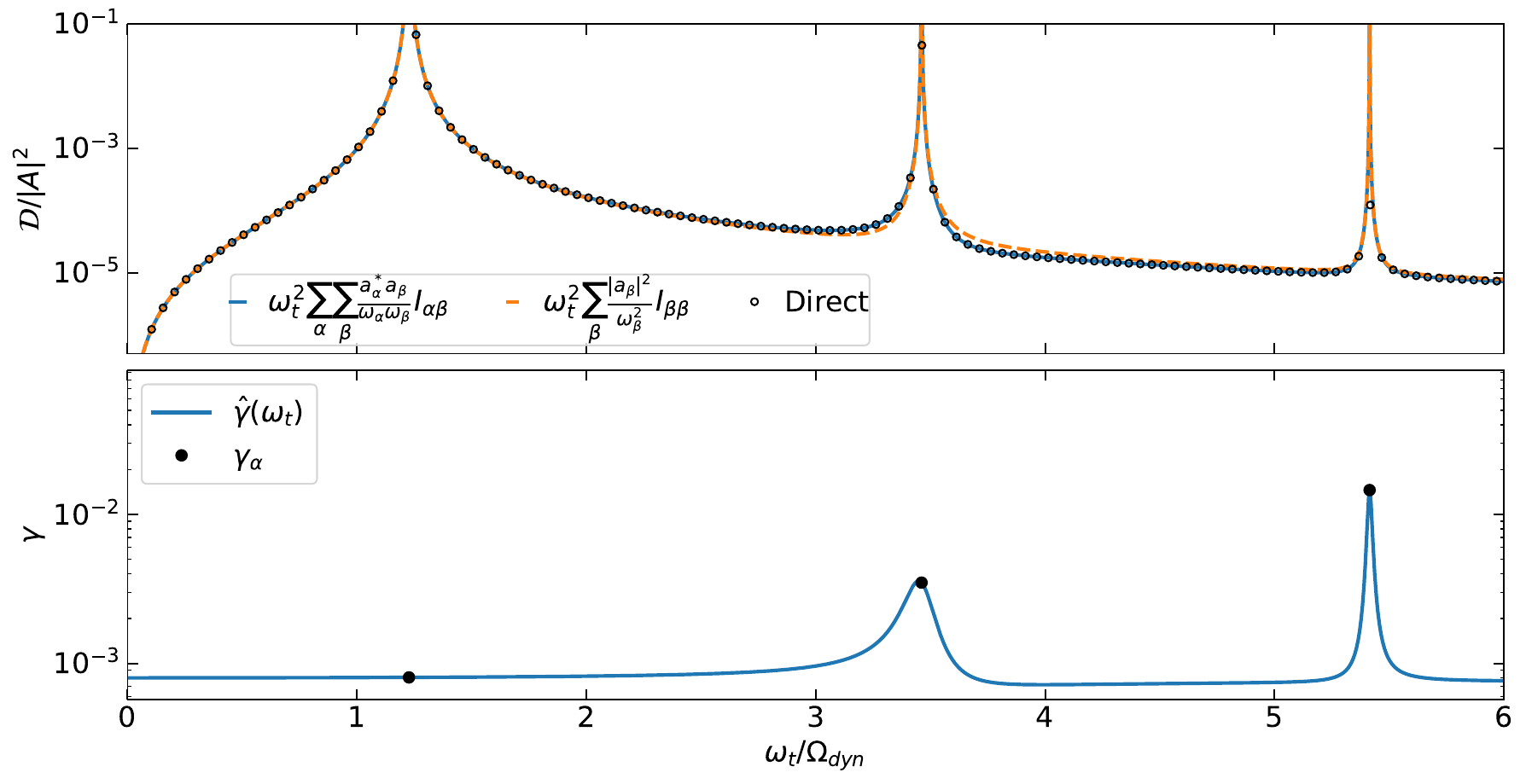}
    \caption{Top: calculations of the time-averaged dissipation rate (per squared amplitude of the tidal potential) as a function of tidal frequency for a non-rotating, $n=1$ polytrope with kinematic viscosity $\nu=10^{-4}R^2\Omega_\text{dyn}.$ The black circles show direct calculations made using the approach described in \citet{Dewberry2023}, which agree extremely well with the results of inserting the adiabatic oscillations and their non-dissipative tidal amplitudes into \autoref{eq:D} (blue curve). On the other hand, the orange curve shows the results of using \autoref{eq:D} with viscous coupling between different modes (i.e., the off-diagonal terms in the left/middle panels of \autoref{fig:int}) excluded. Ignoring the coupling between modes produces discrepant dissipation away from resonance. Bottom: universal but frequency-dependent $\hat{\gamma},$ computed according to \autoref{eq:gmc} (blue curve), as compared with the individual mode damping rates (black dots). The individual mode damping rates apply at resonance, but not away from it.}
    \label{fig:viscnr_gm}
\end{figure*}

Crucially, although the equation of motion only produces a nonzero $\hat{\gamma}$ when dissipation is included, the integral that primarily determines its value (\autoref{eq:D}) is dominated by the structure of the non-dissipative wave response (provided that the dissipation is weak). \autoref{fig:viscnr_gm} (top) demonstrates the agreement between calculations of the dissipation $\mathcal D$ for an isentropic, non-rotating, $n=1$ polytrope with kinematic viscosity $\nu=\mu/\rho_0=10^{-4}R^2\Omega_\text{dyn}$ (where $\Omega_\text{dyn}=(GM/R^3)^{1/2}$ is the dynamical frequency) from an expansion in adiabatic modes whose amplitudes are determined \textit{without} damping (i.e., from \autoref{eq:aafric} but with $\gamma=0$; blue curve), and independent direct solution of the tidal equations at fixed tidal frequencies using the approach of \citet{Dewberry2023} (black dots). This agreement fails when viscous couplings between modes are ignored (orange curve). 

The bottom panel in \autoref{fig:viscnr_gm} plots $\hat{\gamma}(\omega_t)$ computed from this expansion of the dissipation via \autoref{eq:gmc} (blue curve). Considered with purely adiabatic amplitudes, \autoref{eq:gmc} formally breaks down close to resonance with a given mode $\alpha$ (i.e., where $\omega_\alpha=\omega_t$). But at exact resonance the damping rate $\hat{\gamma}$ is given simply by the individual damping rate of the resonant mode (as shown by the black dots). Note that in \autoref{fig:viscnr_gm} we have made no particular effort to enforce the limiting behavior $\hat{\gamma}\rightarrow\gamma_\alpha$ as $\omega_t\rightarrow\omega_\alpha$, simply inserting the values $\hat{\gamma}(\omega_t=\omega_\alpha)=\gamma_\alpha$ at exact resonance. Even under the first-order approximation adopted in deriving it, \autoref{eq:gmc} naturally enforces the appropriate limiting behavior very close to resonance. This follows from the fact that when the summations in \autoref{eq:gmc} come to be  dominated by a single mode $\alpha$, the expression  in fact reduces to $\hat{\gamma}=\gamma_\alpha$. We caution, however, that this happy convergence may fail in situations where the summations are not dominated by a single mode despite proximity to resonance (i.e., close to resonance with a mode that only couples weakly to the tidal force).

\autoref{fig:visc_nonrot} further demonstrates both the inaccuracy of treating each mode's damping independently, and the effectiveness of the approximation given by \autoref{eq:gmc}.
The orange and blue lines show calculations of $\Im[k_{22}]$ computed for a non-rotating, $n=1$ polytrope using \autoref{eq:lovenr} but with $\hat{\gamma}$ replaced by individual mode damping rates $\gamma_\alpha$ (in an expansion including the $\ell=2$ f-mode and the four lowest order p-modes). These $\gamma_\alpha$ have been computed both with the quasi-adiabatic approach (dashed orange line), and also an ``ab initio'' approach of solving for eigenmodes of the viscous linearized equations and simply taking $\gamma_\alpha$ from the imaginary part of the oscillations' frequencies (solid blue line). On the other hand, the green solid line shows the result of using \autoref{eq:lovenr} with $\hat{\gamma}$ determined from \autoref{eq:gmc}. Finally, the open black dots show the results of the direct solution of the viscous tidal equations at the indicated tidal frequencies.

\autoref{fig:visc_nonrot} first of all shows that away from resonance (e.g., near $\omega_t/\Omega_\text{dyn}\simeq3,4$), the mode expansions employing individual damping rates disagree with the direct numerical calculations. This disagreement is small, but significant, reaching relative differences of $25\%-50\%$ close to the resonance at $\omega_t/\Omega_\text{dyn}\simeq 3.5$ (plots showing the real parts of $k_{22}$ are indistinguishable for all the methods used). The profile of $\Im[k_{22}]$ computed with a universal $\hat{\gamma}$ reconciles these differences. Reconstructing the total $\boldsymbol{\xi}$ and $\delta\Phi$ from amplitudes involving our frequency-dependent $\hat{\gamma}$, we also find that our expansion accurately reproduces the total wave response computed through the direct approach, as well as the phase lag of the surface gravitational potential. 

Note that although $\hat{\gamma}$ has been constructed specifically to enforce the correct relationship between $\mathcal{D}$ and $\Im[k_{22}]$, this method of construction remains entirely independent of the direct numerical calculations, since the dissipation is computed from an expansion in the adiabatic oscillation modes. This approach of ``bootstrapping'' therefore improves the mode expansion's accuracy without compromising its usefulness as a sparse representation of the tidal response.

\begin{figure*}
    \centering
    \includegraphics[width=\textwidth]{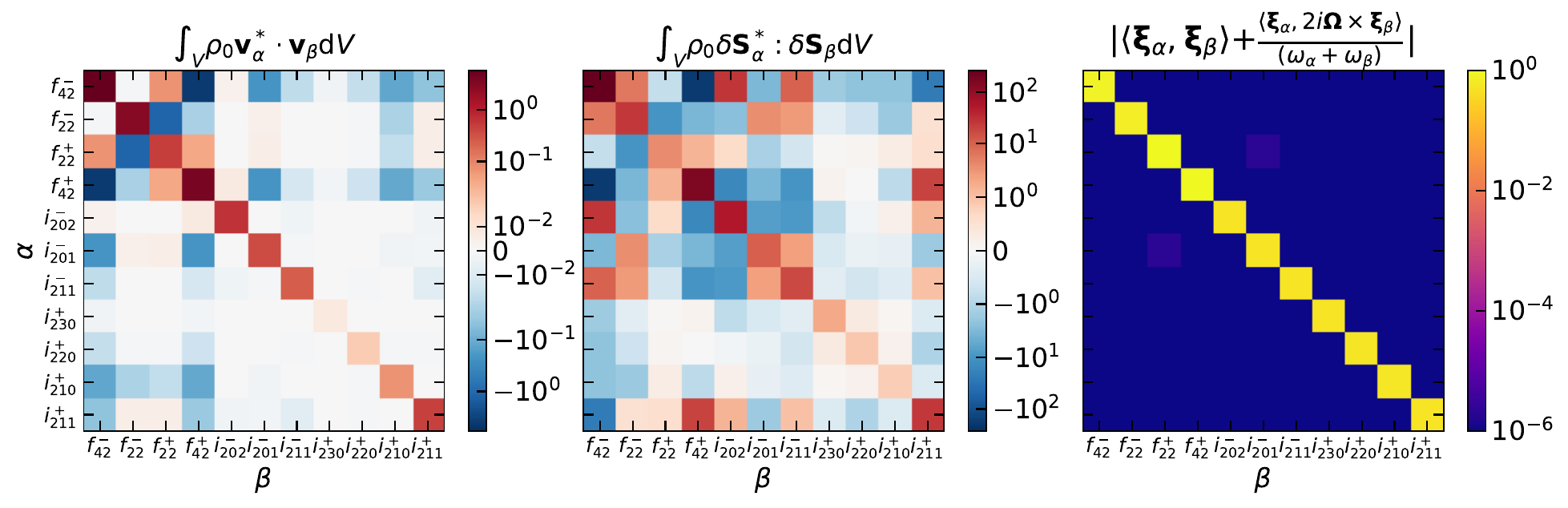}
    \caption{Left, middle: dissipative coupling integrals associated with frictional (left) and viscous (middle) dissipation, computed for the $\ell\simeq2,4$, $m=2$ f-modes and longest wavelength $m=2$ inertial modes of an $n=1$, isentropic polytrope with $\Omega\simeq0.5\Omega_\text{dyn}$. Modes are labelled as in \citet{Dewberry2022a}, and the color-scales transition from log to linear at values $250$ times smaller than the maximum amplitude. Right: modified orthogonality relation \eqref{eq:ortho} computed using the same eigenmodes and quadrature. The orthogonality integrals are less exact than the right panel of \autoref{fig:int} (due to error introduced by integrating in two dimensions), but   still exhibit maximal errors more than five orders of magnitude smaller than the off-diagonal integrals shown in the left and center panels.}
    \label{fig:introt}
\end{figure*}

\section{Rotating bodies}\label{sec:rot}
In this section we generalize our modal calculation of tidal dissipation to accommodate rigid rotation in the tidally perturbed body. Rotation necessitates modification even in the absence of dissipation, since it alters the orthogonality of individual oscillation modes \citep{Schenk2001}. This modified orthogonality similarly complicates the role played by dissipation, even in the simplest case of a frictional damping.

The equation of motion for the Lagrangian displacement $\boldsymbol{\xi}$ induced in a rigidly rotating fluid body with angular velocity ${\bf \Omega}=\Omega\hat{\bf z}$ by a tidal force ${\bf f}$ can be written in the co-rotating frame as
\begin{equation}\label{eq:tEoMr}
    \frac{\partial^2\boldsymbol{\xi}}{\partial t^2}
    +2{\bf \Omega\times}\frac{\partial \boldsymbol{\xi}}{\partial t}
    +{\bf C}[\boldsymbol{\xi}]
    +{\bf D}[\boldsymbol{\xi}]
    ={\bf f}.
\end{equation}
We introduce the phase space expansion 
\begin{equation}
    \left[
        \begin{matrix}
            \boldsymbol{\xi} \\
            \partial_t \boldsymbol{\xi}
        \end{matrix}
    \right]
    =\sum_\beta c_\beta(t)
    \left[
        \begin{matrix}
            \boldsymbol{\xi}_\beta \\
            -\text{i}\omega_\beta\boldsymbol{\xi}_\beta
        \end{matrix}
    \right],
\end{equation}
in unique pairs $(\boldsymbol{\xi}_\beta,\omega_\beta)$ of adiabatic modes satisfying 
\begin{align}
\label{eq:eEoM}
    -\omega_\beta^2\boldsymbol{\xi}_\beta
    -2\text{i}\omega_\beta{\bf \Omega\times}\boldsymbol{\xi}_\beta
    +{\bf C}[\boldsymbol{\xi}_\beta]
    &=0,
\\\label{eq:ortho}
    (\omega_\alpha+\omega_\beta)
    \langle \boldsymbol{\xi}_\alpha,\boldsymbol{\xi}_\beta\rangle
    +\langle 
        \boldsymbol{\xi}_\alpha,
        2\text{i}{\bf \Omega\times}\boldsymbol{\xi}_\beta
    \rangle
    &=0\hspace{1em}\text{for}\hspace{1em}\beta\not=\alpha
\\
    \langle \boldsymbol{\xi}_\beta,\boldsymbol{\xi}_{\beta}\rangle&=MR^2,
\end{align}
where $R$ is now the \emph{equatorial} radius of the rotating body. Throughout the rest of the section, we return to working in units with $M=G=R=1.$
Note that in summation, the phase space expansion implies that
\begin{align}\label{eq:phv}
    \partial_t \boldsymbol{\xi}
    =\sum_\beta \dot{c}_\beta \boldsymbol{\xi}_\beta
    =-\sum_\beta \text{i}\omega_\beta c_\beta \boldsymbol{\xi}_\beta,
\end{align}
although this equality need not hold for individual terms in the series. Substitution into \autoref{eq:tEoM} then produces 
\begin{equation}\label{eq:cEoM}
    \dot{c}_{\alpha}
    +\text{i}\omega_{\alpha}c_{\alpha}
    +\frac{\text{i}}{2\epsilon_\alpha}
    \sum_\beta
    \langle 
        \boldsymbol{\xi}_\alpha,
        {\bf D}[c_\beta\boldsymbol{\xi}_\beta]
    \rangle
    =\frac{\text{i}}{2\epsilon_{\alpha}}\langle \boldsymbol{\xi}_\alpha,-\nabla U \rangle,
\end{equation}
where 
$\epsilon_{\alpha}
=\omega_\alpha
\langle \boldsymbol{\xi}_\alpha,\boldsymbol{\xi}_{\alpha}\rangle
+\langle 
    \boldsymbol{\xi}_\alpha,
    \text{i}{\bf \Omega\times}\boldsymbol{\xi}_{\alpha}
\rangle.$
As in the absence of rotation, the summation on the LHS remains a summation without further information about the dissipative operator ${\bf D}$.

\subsection{Frictional damping}
For a rigidly rotating body with ${\bf v}=\partial_t\boldsymbol{\xi}$ in the rotating frame,\footnote{
For a differentially rotating body, the relationship between ${\bf v}$ and $\boldsymbol{\xi}$ is more complicated.
} the frictional damping 
\begin{equation}
    {\bf D}[\boldsymbol{\xi}]
=2\gamma\partial_t\boldsymbol{\xi}
=2\gamma
\sum_\beta \dot{c}_\beta \boldsymbol{\xi}_\beta
=-2\gamma
\sum_\beta\text{i}\omega_\beta c_\beta\boldsymbol{\xi}_\beta
\end{equation} 
produces
\begin{align}\notag
    \dot{c}_{\alpha}
    &+\text{i}\omega_{\alpha}c_\alpha 
    +\frac{\gamma}{\epsilon_\alpha}
    \left(
        \omega_\alpha c_\alpha
        -\sum_{\beta\not=\alpha} 
        \frac{\omega_\beta c_\beta}{(\omega_\alpha+\omega_\beta)}
        \langle 
            \boldsymbol{\xi}_\alpha,
            2\text{i}\boldsymbol{\Omega}\times 
            \boldsymbol{\xi}_\beta
        \rangle
    \right)
\\& \label{eq:frotceqn}
    =\frac{\text{i}}{2\epsilon_{\alpha}}\langle \boldsymbol{\xi}_\alpha,-\nabla U \rangle.
\end{align}
In the limiting case $\Omega=0,$ $\epsilon_\alpha
=\omega_\alpha\langle \boldsymbol{\xi}_\alpha,
\boldsymbol{\xi}_\alpha\rangle $ and so \autoref{eq:frotceqn} reduces to the decoupled expression
\begin{equation}
    \dot{c}_{\alpha}
    +\text{i}(\omega_{\alpha}-\text{i}\gamma)c_\alpha
    =\frac{\text{i}}{2\epsilon_{\alpha}}\langle \boldsymbol{\xi}_\alpha,-\nabla U \rangle,
\end{equation}
which in a steady state with $\dot{c}_\alpha=-\text{i}\omega_t c_\alpha$ gives
\begin{equation}\label{eq:cafric}
    c_\alpha
    =\frac{-\langle \boldsymbol{\xi}_\alpha,\nabla U \rangle}
    {2\epsilon_{\alpha}(\omega_{\alpha}-\omega_t-\text{i}\gamma)}.
\end{equation}
For nonzero rotation, however, even the simple case of a frictional damping does not produce decoupled amplitude equations, due to the modified orthogonality relation \eqref{eq:ortho}. 

Like \autoref{fig:int}, \autoref{fig:introt} demonstrates the coupling across the oscillation modes of an $n=1,$ isentropic polytrope with $\Omega\simeq0.5\Omega_\text{dyn}$. The left (middle) panels show integrals proportional to 
$\langle \boldsymbol{\xi}_\alpha,
{\bf D}[\boldsymbol{\xi}_\beta]\rangle$ 
for frictional (viscous) damping, computed for the $m=2,$ $\ell\simeq2,4$ f-modes and the seven (three retrograde and four prograde) longest wavelength inertial modes. The axis labels adopt the same naming scheme as \citet{Dewberry2022a}. Frictional damping provides the strongest coupling between the prograde/retrograde f-modes of the same degree \citep[as found in viscous Maclaurin spheroids by ][]{Braviner2015}, and between the $\ell\simeq4$ f-modes and the two longest wavelength inertial modes ($i_{201}^-$ and $i_{210}^+$). Viscosity provides similarly strong coupling between the same modes, but also among the rest of the oscillations. Both forms of damping involve significant ``off-diagonal'' coupling between different modes. 

\begin{figure*}
    \centering
    \includegraphics[width=\textwidth]{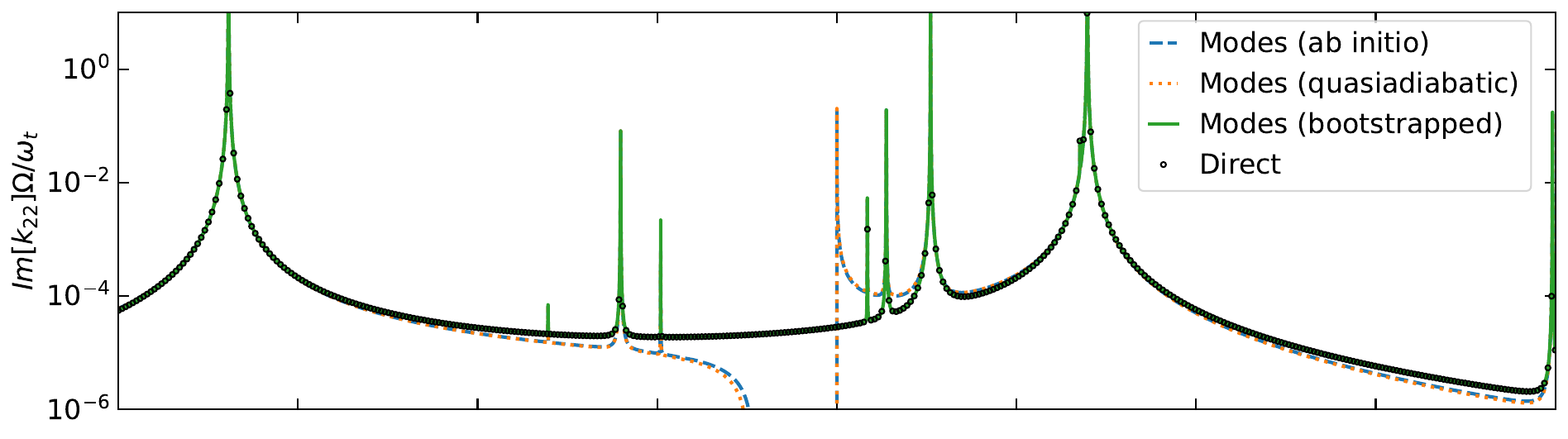}
    \includegraphics[width=\textwidth]{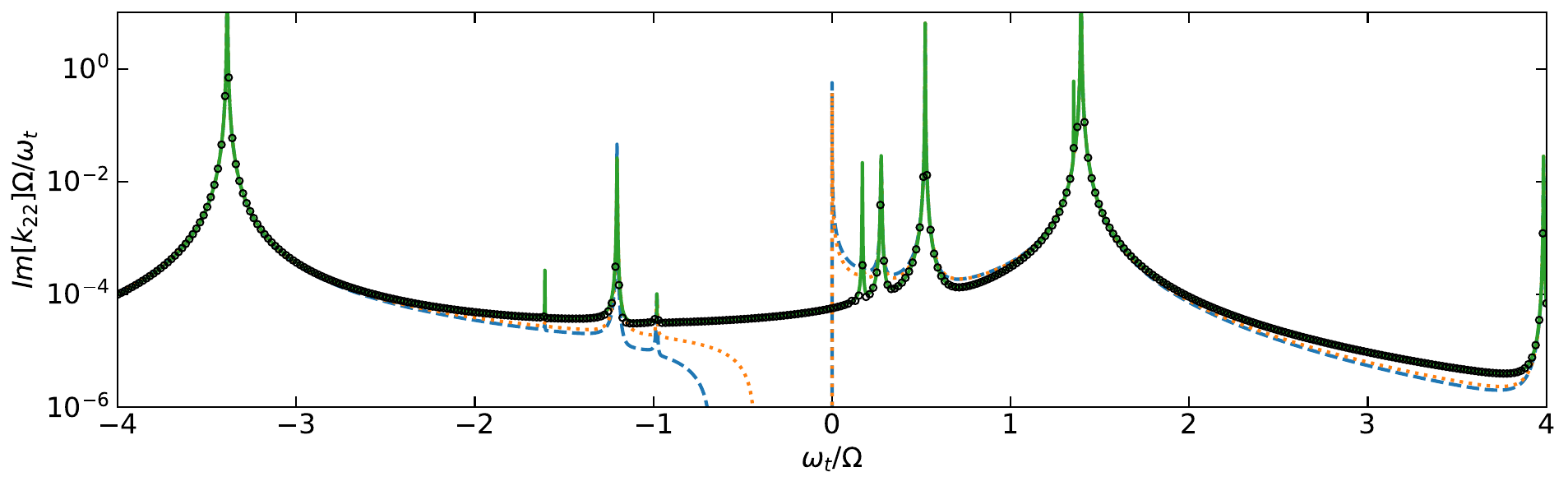}
    \caption{Same as \autoref{fig:visc_nonrot}, but for an $n=1$ rotating polytrope with $\Omega=0.5\Omega_\text{dyn}$ and both frictional (top, $\gamma=0.5\times10^{-4}\Omega_\text{dyn})$ and viscous (bottom,  $\nu=10^{-5}R^2\Omega_\text{dyn}$) dissipation. Division of $\Im[k_{22}]$ by $\omega_t$ highlights the fact that the expansions adopting individual mode damping rates produce tidal torques with the wrong sign in the regime $-\Omega\lesssim\omega_t\lesssim0$. Our approach of bootstrapping a universal damping rate $\hat{\gamma}$ avoids this error.}
    \label{fig:rot}
\end{figure*}

\subsection{Effective damping rate for rotating bodies}
A similar procedure to that described in \autoref{sec:nonrot} can be used to construct a universal, frequency-dependent $\hat{\gamma}(\omega_t).$ Assuming that the collective frictional damping of the tidally driven oscillations produces a frequency-dependent $\hat{\gamma}(\omega_t)$ such that
\begin{equation}
    \sum_\beta\langle 
        \boldsymbol{\xi}_\alpha,
        {\bf D}[c_\beta\boldsymbol{\xi}_\beta]
    \rangle
    =-2\text{i}\hat{\gamma}\epsilon_\alpha c_\alpha,
\end{equation}
Love numbers can be computed from a phase space expansion in tidally driven modes via
\begin{equation}
    k_{\ell m}
    =\frac{2\pi}{(2\ell+1)}
    \sum_{\alpha}
    \frac{|Q_{\ell m}^\alpha|^2}
    {\epsilon_{\alpha}(\omega_\alpha-\omega_t - \text{i}\hat{\gamma})}.
\end{equation}
The time-averaged dissipation due to a frictional damping ${\bf D}[\boldsymbol{\xi}]=2\gamma{\bf v}$ is given by
\begin{equation}
    \mathcal{D}=\int_V\gamma \rho_0|{\bf v}|^2\text{d}V,
\end{equation}
and the tidal velocity field can be written in terms of non-dissipative amplitudes as
\begin{equation}
    {\bf v}=
    \sum_\alpha c_\alpha {\bf v}_\alpha
    =-\sum_\alpha \frac{\langle \boldsymbol{\xi}_\alpha,\nabla U\rangle}
    {2\epsilon_{\alpha}(\omega_\alpha-\omega_t)}
    \left(\frac{\omega_t}{\omega_\alpha}\right){\bf v}_\alpha.
\end{equation}
The factor of $\omega_t/\omega_\alpha$ is consistent, but not formally necessary with use of a phase space expansion (given \autoref{eq:phv}). When the expansion is truncated, though, we find that this factor is essential for ensuring that the dissipation rate vanishes as $\omega_t\rightarrow0.$ 

Relating the dissipation to the imaginary part of the tidal Love numbers then yields
\begin{align}\label{eq:Dgmrot}
    \frac{\mathcal{D}}{|A|^2}
    &=\frac{\omega_t}{4}\sum_{\alpha}
        \frac{|Q_{\ell m}^\alpha|^2\hat{\gamma}}
        {\epsilon_{\alpha}(\Delta\omega_\alpha^2 + \hat{\gamma}^2)}
\\\notag&
    =\frac{\omega_t}{4}\sum_{\alpha}
    \frac{|Q_{\ell m}^\alpha|^2}
    {\epsilon_{\alpha}\Delta\omega_\alpha}
    \left[
        \frac{\hat{\gamma}}{\Delta\omega_\alpha} - \frac{\hat{\gamma}^3}{\Delta\omega_\alpha^3}
        +\mathcal O\left(\frac{\hat{\gamma}^5}{\Delta\omega_\alpha^5}\right)
    \right],
\end{align}
where $\Delta\omega_\alpha=\omega_\alpha - \omega_t.$ To leading order in $\hat{\gamma}/\Delta\omega_\alpha$ (which again is small except near exact resonance), we then find
\begin{align}\label{eq:gmfricrot}
    \hat{\gamma}
    &\simeq\frac{4}{\omega_t}\left(
        \sum_{\alpha}
        \frac{|Q_{\ell m}^\alpha|^2}
        {\epsilon_{\alpha}\Delta\omega_\alpha^2}
    \right)^{-1}\frac{\mathcal{D}}{|A|^2}
\\\notag&
    =\left(
        \sum_{\alpha}
        \frac{|Q_{\ell m}^\alpha|^2}
        {\epsilon_{\alpha}\Delta\omega_\alpha^2}
    \right)^{-1}
    \sum_{\alpha,\beta}
        \frac{4\omega_tQ_{\ell m}^{\alpha}Q_{\ell m}^\beta}
        {E_{\alpha}E_\beta
        \Delta\omega_\alpha\Delta\omega_\beta}
        \int_V
        \gamma \rho_0{\bf v}_\alpha^*\cdot {\bf v}_\beta\text{d}V
    ,
\end{align}
where $E_\alpha=2\omega_\alpha\epsilon_\alpha$ is the mode energy at unit amplitude in the rotating frame \citep{Schenk2001}. 
In the case of viscous dissipation, the same argument leads to 
\begin{equation}\label{eq:gmviscrot}
    \hat{\gamma}
    \simeq
    \left(
        \sum_{\alpha}
        \frac{|Q_{\ell m}^\alpha|^2}
        {\epsilon_{\alpha}\Delta\omega_\alpha^2}
    \right)^{-1}
    \sum_{\alpha,\beta}
    \frac{4\omega_tQ_{\ell m}^{\alpha}Q_{\ell m}^\beta}
    {E_{\alpha}E_\beta\Delta\omega_\alpha\Delta\omega_\beta}
    I_{\alpha\beta}.
\end{equation}

\begin{figure*}
    \centering
    \includegraphics[width=\textwidth]{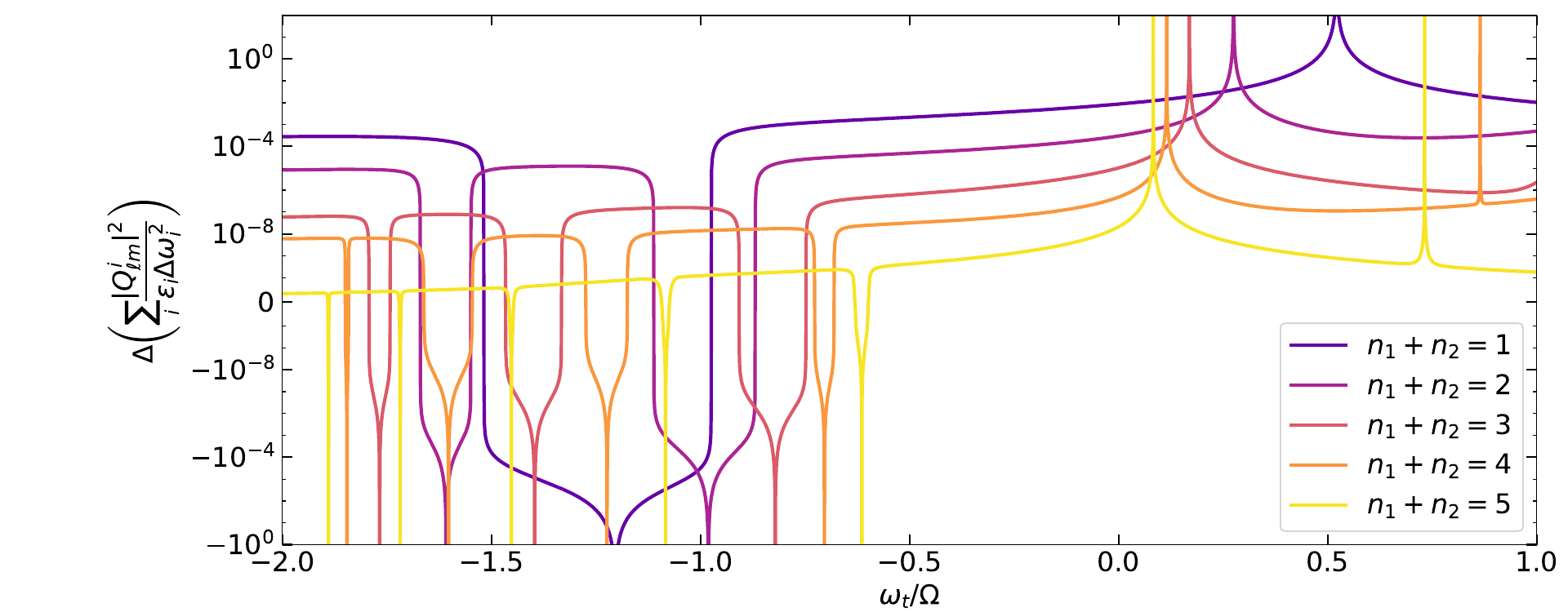}
    \caption{Contributions to the summation in \autoref{eq:k22exp} due to inertial modes with successively shorter wavelengths \citep[characterized by ``vertical'' and ``radial'' quantum numbers; ][]{Wu2005a}. The retrograde inertial modes produce positive contributions except near exact resonance, implying that a vary large quantity of short-wavelength modes would be needed to drive $\Im[k_{22}]$ to negative values where $\omega_t<0$ (as required for a stable body).}
    \label{fig:imodeup}
\end{figure*}

The top and bottom panels of \autoref{fig:rot} respectively demonstrate the efficacy of \autoref{eq:gmfricrot} and \autoref{eq:gmviscrot} for frictional and viscous damping in the $n=1,$ isentropic polytrope with rotation rate $\Omega\simeq 0.5\Omega_\text{dyn}$. Notably, the mode expansions adopting individual mode damping rates\footnote{
Computed with the quasi-adiabatic approach as $\gamma_\alpha=\int_V\gamma \rho_0|{\bf v}_\alpha|^2\text{d}V/(\omega_\alpha\epsilon_\alpha)$ and $I_{\alpha\alpha}/(\omega_\alpha\epsilon_\alpha)$ for frictional and viscous dissipation (respectively).
} even produce imaginary parts of Love numbers that have the wrong sign in the frequency range $-\Omega\lesssim\omega_t\lesssim0$ \citep[positive definite dissipation requires that Love numbers' imaginary parts have the same sign as the tidal frequency in the rotating frame; ][]{Ogilvie2013}. This is in part due to the use of a truncated expansion in inertial modes, which densely populate the frequency range $|\omega|<2\Omega$. To see why, note that near $|\omega_t|\simeq0$,
\begin{align}\label{eq:k22exp}
    \Im[k_{\ell m}]
    &\simeq \frac{2\pi\hat{\gamma}}{2\ell+1}\sum_\beta 
    \frac{|Q_{\ell m}^\beta|^2}{\epsilon_\beta\Delta\omega_\beta^2}
\\\notag&
    \simeq \frac{2\pi\hat{\gamma}}{2\ell+1}
    \left(
        \frac{|Q_{\ell m}^{f-}|^2}{\epsilon_{f-}\Delta\omega_{f-}^2}
        +\frac{|Q_{\ell m}^{f+}|^2}{\epsilon_{f+}\Delta\omega_{f+}^2}
        +\sum_i\frac{|Q_{\ell m}^i|^2}{\epsilon_i\Delta\omega_i^2}
    \right),
\end{align}
where the subscripts $f-$ and $f+$ denote the retrograde and prograde f-modes dominated by degree $\ell$ (which are responsible for the largest peaks in \autoref{fig:rot}), and subscripts $i$ denote inertial modes. The overall sign of the summation in parentheses in \autoref{eq:k22exp} is determined by the signs of the coefficients $\epsilon_\alpha$, which are generically positive (negative) for prograde (retrograde) modes. The f-mode contributions do not exactly balance in the rotating model, and so inertial mode contributions are required to ensure that $\Im[k_{\ell m}]$ changes sign at $\omega_t=0.$ 

\autoref{fig:imodeup} plots changes in the profile of $\sum_\beta|Q_{\ell m}^\beta|^2/(\epsilon_\beta\Delta\omega_\beta^2)$ as more and more inertial modes of the rotating $n=1$ polytrope are included in the summation. Each curve describes the change associated with the inclusion of successively shorter wavelength inertial modes with the indicated values of $n_1+n_2,$ where $n_1$ and $n_2$ are the quantum numbers described in \cite{Wu2005a} and \cite{Dewberry2022a}. In order to drive $\Im[k_{\ell m}]$ to negative values in the region $-\Omega<\omega_t<0,$ we expect that these curves must also become negative at the same frequencies. However, \autoref{fig:imodeup} shows that at a given value of $n_1+n_2$, the retrograde inertial modes lead to negative shifts in the summation only very close to resonance. Everywhere else, the shift is dominated by the small but positive contributions from the lowest frequency prograde inertial modes with resonances between $0\lesssim\omega_t/\Omega\lesssim0.5$.\footnote{As found by \citet{Dewberry2022a}, these prograde inertial modes gain a larger gravitational influence in rapidly rotating models due to rotational mixing with the prograde sectoral f-mode.} 

\begin{figure*}
    \centering
    \includegraphics[width=\textwidth]{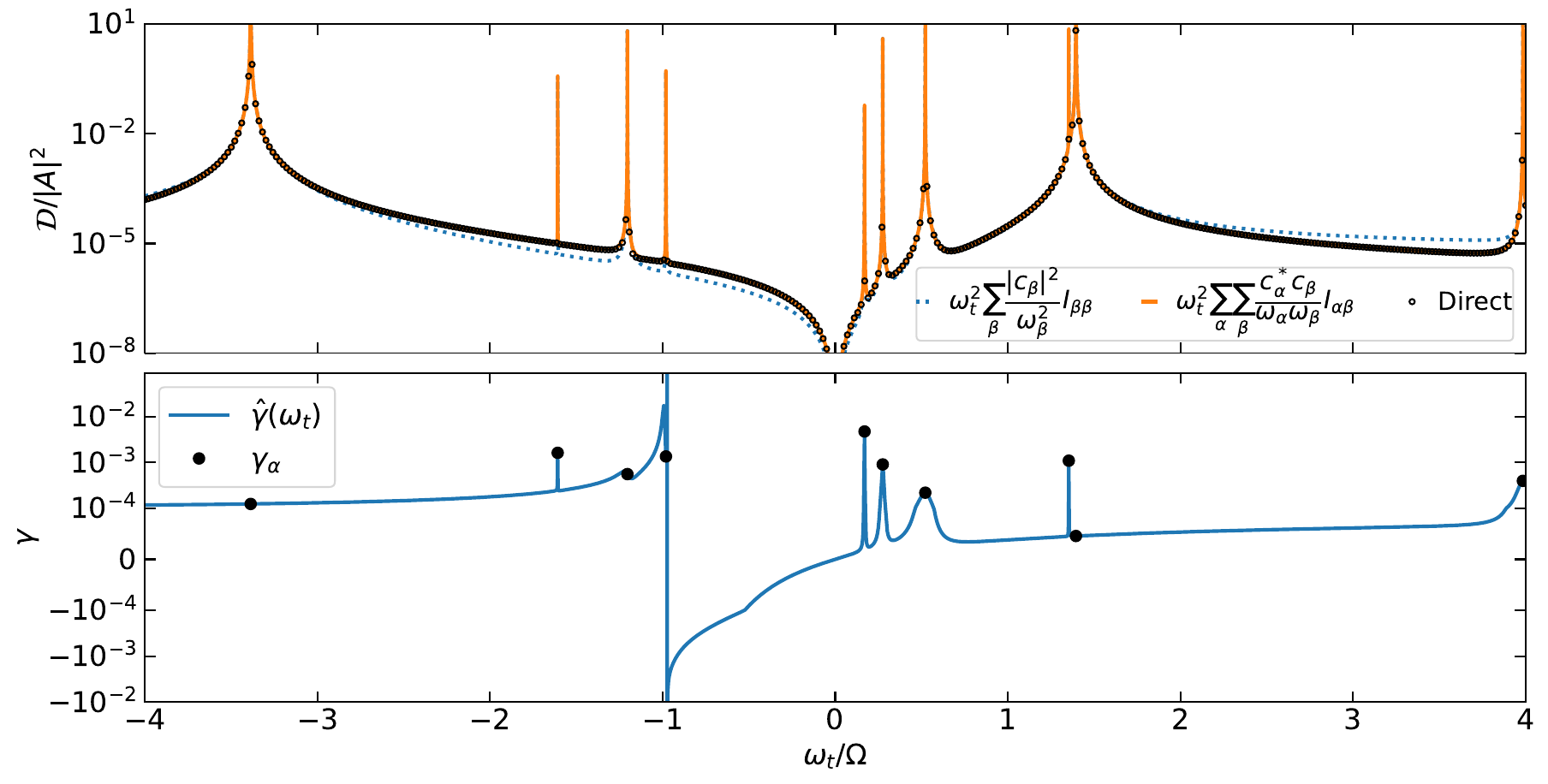}
    \caption{Same as \autoref{fig:viscnr_gm}, but for an $n=1$, isentropic polytrope with $\Omega\simeq0.5\Omega_\text{dyn}$. A truncated expansion in inertial modes (at $n_1+n_2=2$) leads to dissipation rates that are accurate (except near $|\omega_t|\simeq=0$), but also produces effective damping rates $\hat{\gamma}$ that are negative in the range $-\Omega\lesssim\omega_t\lesssim0$.}
    \label{fig:Om5Dg}
\end{figure*}

Consequently, \autoref{fig:imodeup} suggests that ensuring $\sum_\beta|Q_{\ell m}^\beta|^2/(\epsilon_\beta\Delta\omega_\beta^2)<0$ in the frequency range $-\Omega\lesssim\omega_t\lesssim0$ would require the inclusion of an impractical number of inertial modes. On the other hand, only the longest wavelength inertial modes produce resonances large enough to have any discernible impact on $\Im[k_{22}]$ and $\mathcal{D}$ (see \autoref{fig:rot} and \autoref{fig:Om5Dg}, top). Computing $\hat{\gamma}$ from the time-averaged dissipation (via \autoref{eq:gmfricrot} or \autoref{eq:gmviscrot}) therefore permits an efficient recovery of the appropriate relationship between the dissipation and the imaginary part of the Love number. The caveat is that, as the only other quantity in \autoref{eq:k22exp} available to change the sign of $\Im[k_{\ell m}],$ this approach yields negative values of $\hat{\gamma}<0$ in the frequency range $-\Omega\lesssim\omega_t\lesssim0$ (see \autoref{fig:Om5Dg}, bottom). A negative $\hat{\gamma}$ is not physical for a stable body, but we argue that this is immaterial so long as (i) the dissipation rate $\mathcal{D}$ remains unaffected by the exclusion of shorter wavelength modes, and (ii) $\Im[k_{22}]$ satisfies \autoref{eq:diss}. The latter point holds by construction, and we find that the former only requires truncation at $n_1+n_2\simeq2$ even for the rapidly rotating model considered here.

An additional complication is that the transition to negative $\hat{\gamma}$ at $\omega_t/\Omega\simeq-1$ leads to a breakdown in the assumption that $\hat{\gamma}/\Delta\omega_\alpha$ is small (at least for the truncated mode expansion used here). This breakdown can in turn produce spurious values of $\Im[k_{22}]$ with the wrong sign when only the leading order terms in $\hat{\gamma}/\Delta\omega_\alpha$ are retained in \autoref{eq:Dgmrot}. These spurious values can be corrected by retaining higher-order terms when computing $\hat{\gamma}$, however. At all other frequencies the use of \autoref{eq:gmfricrot} or \autoref{eq:gmviscrot} with a relatively sparse mode expansion (involving four f-modes and seven inertial modes) does an excellent job of reproducing the dissipation rates and Love numbers found through direct calculations.

We do find that truncation of the mode expansion can lead to a less accurate reconstruction of the total wave response (as measured by a comparison of, e.g., $\boldsymbol{\xi}$ computed with mode expansions vs. the direct approach) than in the nonrotating case, depending on tidal frequency. However, this holds true for treatments involving individual damping rates as well. Since tidally interacting bodies communicate through gravity, accurate calculations of potential Love numbers are much more relevant to the prediction of secular tidal evolution.

\section{Conclusions}\label{sec:conc}
Oscillation mode expansions provide a useful and often enlightening framework for describing the response of a star or gaseous planet to tidal perturbation. However, \citet{Townsend2023} demonstrate that the tidal torques computed via mode expansions involving g-modes individually damped by radiative diffusion deviate from those derived via direct solution of the governing partial differential equations. The problem lies with the application of decoupled driven/damped harmonic oscillator equations to tidal solutions that may not satisfy them, depending on the form of dissipation included in the equation of motion \citep[e.g.,][]{Braviner2015}. 

We have expanded upon the analyses of \citet{Townsend2023}, and confirmed similar discrepancies between modal and direct solutions involving frictional and viscous dissipation. Moreover, we have demonstrated that expansions involving only the adiabatic oscillation modes can still reproduce the results of (more numerically expensive) direct calculations, but only if wave damping is treated collectively. 

In non-rotating and rigidly rotating bodies, we find that fundamental relationships between energy dissipation and the imaginary parts of tidal Love numbers \citep{Ogilvie2013} permit the construction of universal (but frequency-dependent) damping rates that can be applied in the usual \citep[e.g.,][]{Schenk2001} decoupled amplitude equations. Our approach bypasses the need for the inclusion of a dense spectrum of short-wavelength inertial modes that would otherwise complicate mode expansions for fully convective bodies; we find that tidal torques can be computed accurately at most relevant tidal frequencies with only a modest set of modes, even for a model rotating at half the break-up angular velocity.

In this work we have limited our focus to simple (polytropic) models with simple dissipation (constant frictional damping or constant kinematic viscosity). However, the fundamental nature of the relationship between the dissipation rate and the imaginary parts of tidal Love numbers implies that this approach should hold for, e.g., stars affected by radiative damping. In a companion paper \citep{Wu2023}, we explore the tidal torques of more realistic stellar models.

\section*{Acknowledgements}
We thank the anonymous referee for helpful comments that improved the quality of the paper. J. W. D. is supported by the Natural
Sciences and Engineering Research Council of Canada (NSERC), [funding
reference \#CITA 490888-16]. Through S.C.W., this material is based upon work supported by the National Science Foundation Graduate Research Fellowship under Grant No. DGE‐1745301.

\section*{Data Availability}
The data underlying this work will be provided upon reasonable
request to the corresponding author.



\bibliographystyle{mnras}
\bibliography{modamp} 

\begin{thebibliography}{}
\makeatletter
\relax
\def\mn@urlcharsother{\let\do\@makeother \do\$\do\&\do\#\do\^\do\_\do\%\do\~}
\def\mn@doi{\begingroup\mn@urlcharsother \@ifnextchar [ {\mn@doi@}
  {\mn@doi@[]}}
\def\mn@doi@[#1]#2{\def\@tempa{#1}\ifx\@tempa\@empty \href
  {http://dx.doi.org/#2} {doi:#2}\else \href {http://dx.doi.org/#2} {#1}\fi
  \endgroup}
\def\mn@eprint#1#2{\mn@eprint@#1:#2::\@nil}
\def\mn@eprint@arXiv#1{\href {http://arxiv.org/abs/#1} {{\tt arXiv:#1}}}
\def\mn@eprint@dblp#1{\href {http://dblp.uni-trier.de/rec/bibtex/#1.xml}
  {dblp:#1}}
\def\mn@eprint@#1:#2:#3:#4\@nil{\def\@tempa {#1}\def\@tempb {#2}\def\@tempc
  {#3}\ifx \@tempc \@empty \let \@tempc \@tempb \let \@tempb \@tempa \fi \ifx
  \@tempb \@empty \def\@tempb {arXiv}\fi \@ifundefined
  {mn@eprint@\@tempb}{\@tempb:\@tempc}{\expandafter \expandafter \csname
  mn@eprint@\@tempb\endcsname \expandafter{\@tempc}}}

\bibitem[\protect\citeauthoryear{Batchelor}{Batchelor}{2000}]{Batchelor2000}
Batchelor G.~K.,  2000, An Introduction to Fluid Dynamics.
Cambridge Mathematical Library, Cambridge University Press,
  \mn@doi{10.1017/CBO9780511800955}

\bibitem[\protect\citeauthoryear{{Braviner} \& {Ogilvie}}{{Braviner} \&
  {Ogilvie}}{2015}]{Braviner2015}
{Braviner} H.~J.,  {Ogilvie} G.~I.,  2015, \mn@doi [\mnras]
  {10.1093/mnras/stu2521}, \href
  {https://ui.adsabs.harvard.edu/abs/2015MNRAS.447.1141B} {447, 1141}

\bibitem[\protect\citeauthoryear{{Burkart}, {Quataert}, {Arras}  \&
  {Weinberg}}{{Burkart} et~al.}{2012}]{Burkart2012}
{Burkart} J.,  {Quataert} E.,  {Arras} P.,   {Weinberg} N.~N.,  2012, \mn@doi
  [\mnras] {10.1111/j.1365-2966.2011.20344.x}, \href
  {https://ui.adsabs.harvard.edu/abs/2012MNRAS.421..983B} {421, 983}

\bibitem[\protect\citeauthoryear{{Dewberry}}{{Dewberry}}{2023}]{Dewberry2023}
{Dewberry} J.~W.,  2023, \mn@doi [\mnras] {10.1093/mnras/stad546}, \href
  {https://ui.adsabs.harvard.edu/abs/2023MNRAS.521.5991D} {521, 5991}

\bibitem[\protect\citeauthoryear{{Dewberry} \& {Lai}}{{Dewberry} \&
  {Lai}}{2022}]{Dewberry2022a}
{Dewberry} J.~W.,  {Lai} D.,  2022, \mn@doi [\apj] {10.3847/1538-4357/ac3ede},
  \href {https://ui.adsabs.harvard.edu/abs/2022ApJ...925..124D} {925, 124}

\bibitem[\protect\citeauthoryear{{Fuller}}{{Fuller}}{2017}]{Fuller2017}
{Fuller} J.,  2017, \mn@doi [\mnras] {10.1093/mnras/stx2135}, \href
  {https://ui.adsabs.harvard.edu/abs/2017MNRAS.472.1538F} {472, 1538}

\bibitem[\protect\citeauthoryear{{Ipser} \& {Lindblom}}{{Ipser} \&
  {Lindblom}}{1991}]{Ipser1991}
{Ipser} J.~R.,  {Lindblom} L.,  1991, \mn@doi [\apj] {10.1086/170039}, \href
  {https://ui.adsabs.harvard.edu/abs/1991ApJ...373..213I} {373, 213}

\bibitem[\protect\citeauthoryear{{Kumar}, {Ao}  \& {Quataert}}{{Kumar}
  et~al.}{1995}]{Kumar1995}
{Kumar} P.,  {Ao} C.~O.,   {Quataert} E.~J.,  1995, \mn@doi [\apj]
  {10.1086/176055}, \href
  {https://ui.adsabs.harvard.edu/abs/1995ApJ...449..294K} {449, 294}

\bibitem[\protect\citeauthoryear{{Ogilvie}}{{Ogilvie}}{2009}]{Ogilvie2009}
{Ogilvie} G.~I.,  2009, \mn@doi [\mnras] {10.1111/j.1365-2966.2009.14814.x},
  \href {https://ui.adsabs.harvard.edu/abs/2009MNRAS.396..794O} {396, 794}

\bibitem[\protect\citeauthoryear{{Ogilvie}}{{Ogilvie}}{2013}]{Ogilvie2013}
{Ogilvie} G.~I.,  2013, \mn@doi [\mnras] {10.1093/mnras/sts362}, \href
  {https://ui.adsabs.harvard.edu/abs/2013MNRAS.429..613O} {429, 613}

\bibitem[\protect\citeauthoryear{{Ogilvie}}{{Ogilvie}}{2014}]{Ogilvie2014}
{Ogilvie} G.~I.,  2014, \mn@doi [\araa] {10.1146/annurev-astro-081913-035941},
  \href {https://ui.adsabs.harvard.edu/abs/2014ARA&A..52..171O} {52, 171}

\bibitem[\protect\citeauthoryear{{Press} \& {Teukolsky}}{{Press} \&
  {Teukolsky}}{1977}]{Press1977}
{Press} W.~H.,  {Teukolsky} S.~A.,  1977, \mn@doi [\apj] {10.1086/155143},
  \href {https://ui.adsabs.harvard.edu/abs/1977ApJ...213..183P} {213, 183}

\bibitem[\protect\citeauthoryear{{Schenk}, {Arras}, {Flanagan}, {Teukolsky}  \&
  {Wasserman}}{{Schenk} et~al.}{2001}]{Schenk2001}
{Schenk} A.~K.,  {Arras} P.,  {Flanagan} {\'E}.~{\'E}.,  {Teukolsky} S.~A.,
  {Wasserman} I.,  2001, \mn@doi [\prd] {10.1103/PhysRevD.65.024001}, \href
  {https://ui.adsabs.harvard.edu/abs/2001PhRvD..65b4001S} {65, 024001}

\bibitem[\protect\citeauthoryear{{Sun}, {Townsend}  \& {Guo}}{{Sun}
  et~al.}{2023}]{Sun2023}
{Sun} M.,  {Townsend} R.~H.~D.,   {Guo} Z.,  2023, \mn@doi [\apj]
  {10.3847/1538-4357/acb33a}, \href
  {https://ui.adsabs.harvard.edu/abs/2023ApJ...945...43S} {945, 43}

\bibitem[\protect\citeauthoryear{{Townsend} \& {Sun}}{{Townsend} \&
  {Sun}}{2023}]{Townsend2023}
{Townsend} R.~H.~D.,  {Sun} M.,  2023, \mn@doi [arXiv e-prints]
  {10.48550/arXiv.2306.06429}, \href
  {https://ui.adsabs.harvard.edu/abs/2023arXiv230606429T} {p. arXiv:2306.06429}

\bibitem[\protect\citeauthoryear{{Wu}}{{Wu}}{2005a}]{Wu2005a}
{Wu} Y.,  2005a, \mn@doi [\apj] {10.1086/497354}, \href
  {https://ui.adsabs.harvard.edu/abs/2005ApJ...635..674W} {635, 674}

\bibitem[\protect\citeauthoryear{{Wu}}{{Wu}}{2005b}]{Wu2005b}
{Wu} Y.,  2005b, \mn@doi [\apj] {10.1086/497355}, \href
  {https://ui.adsabs.harvard.edu/abs/2005ApJ...635..688W} {635, 688}

\bibitem[\protect\citeauthoryear{{Wu}, {Dewberry}  \& {Fuller}}{{Wu}
  et~al.}{2023}]{Wu2023}
{Wu} S.~C.,  {Dewberry} J.~W.,   {Fuller} J.,  2023, \apj, in prep.

\bibitem[\protect\citeauthoryear{{Xu} \& {Lai}}{{Xu} \& {Lai}}{2017}]{Xu2017}
{Xu} W.,  {Lai} D.,  2017, \mn@doi [\prd] {10.1103/PhysRevD.96.083005}, \href
  {https://ui.adsabs.harvard.edu/abs/2017PhRvD..96h3005X} {96, 083005}

\bibitem[\protect\citeauthoryear{{Yu}, {Weinberg}  \& {Arras}}{{Yu}
  et~al.}{2021}]{Yu2021}
{Yu} H.,  {Weinberg} N.~N.,   {Arras} P.,  2021, \mn@doi [\apj]
  {10.3847/1538-4357/ac0a79}, \href
  {https://ui.adsabs.harvard.edu/abs/2021ApJ...917...31Y} {917, 31}

\makeatother
\end{thebibliography}






\bsp	
\label{lastpage}
\end{document}